# Evolution of defect signatures at ferroelectric domain walls in Mg-doped LiNbO$_3$

Guillaume F. Nataf[*,1,2], Mael Guennou[**,1], Alexander Haußmann[3], Nick Barrett[2] and Jens Kreisel[1,4]

[1] Department Material Research and Technology, Luxembourg Institute of Science and Technology, 41 Rue du Brill, L-4422 Belvaux, Luxembourg
[2] Service de Physique de l'Etat Condensé, DSM/IRAMIS/SPEC, CNRS UMR 3680, CEA Saclay F-91191 Gif sur Yvette cedex, France
[3] Institut für Angewandte Photophysik, Technische Universität Dresden, George-Bähr-Str. 1, D-01069 Dresden, Germany
[4] Physics and Materials Science Research Unit, University of Luxembourg, 41 Rue du Brill, L-4422 Belvaux, Luxembourg

---

[*] Corresponding author: e-mail guillaume.nataf@list.lu, Phone: +352 470 261 513
[**] e-mail mael.guennou@list.lu, Phone: +352 470 261 512

**Abstract**

The domain structure of uniaxial ferroelectric lithium niobate single crystals is investigated using Raman spectroscopy mapping. The influence of doping with magnesium and poling at room temperature is studied by analysing frequency shifts at domain walls and their variations with dopant concentration and annealing conditions. It is shown that defects are stabilized at domain walls and that changes in the defect structures with Mg concentration can be probed by the shift of Raman modes. We show that the signatures of polar defects in the bulk and at the domain walls differ.

## 1. Introduction

Since the 80's, studies of the interaction of charge defects and ferroelectric domain walls have focused on pinning-depinning mechanisms [1-3] and their influence on polarisation fatigue [2,3], a major issue in ferroelectric non-volatile memories. More recently, this topic has received a renewed interest due to the emerging field of so-called domain wall nanoelectronics [4] or domain boundary engineering [5], which hold the promise of using the distinct functional properties of domain walls in electronic devices. Within this context, the crucial role of defects in the electrical conductivity of domain walls has been amply pointed out, even though the precise mechanisms are still debated. For instance, in the case of multiferroic $BiFeO_3$, defects and specifically oxygen vacancies are proposed to play an important role in the conduction at the walls, as compared to the relatively small band gap reduction [6-8]. In $Pb(Zr,Ti)O_3$ thin films, it has been argued that defect segregation could explain conduction at domain walls via hopping between trapped states [9]. Recent reports of photoconductivity at ferroelectric domain walls in single crystals of lithium niobate [10,11] also call for a deeper understanding, in relation to the other phenomena observed at domain walls such as index contrast [12], long-range strain [13,14], phonon spectra changes [15-23] etc. In addition, the increasing demand of ferroelectric patterns with sub-micron domain sizes for photonic applications [24,25] requires a deeper understanding of the impact of polar defects [24].

Here, we investigate interactions between domain walls and defects in lithium niobate ($LiNbO_3$). $LiNbO_3$ is a uniaxial ferroelectric with a rhombohedral structure (*R3c*) at room temperature. Two types of ferroelectric domains are permitted; they are related by the loss of inversion symmetry at the phase transition, and differ only by their polarization orientation. Generally speaking, polar defects may align along or against the ferroelectric polarization, and break this equivalence. The presence of defects in $LiNbO_3$ is particularly crucial in its congruent composition, i.e. when the ratio C = Li/(Li+Nb) is equal to 0.485, and it is likely that in this situation defects strongly interact with domain walls. Three main models have been proposed for point defects and charge compensation in lithium niobate involving: (i) lithium vacancies and oxygen vacancies ($2V'_{Li} + V^{\cdot\cdot}_O$) [26], (ii) niobium antisites and niobium vacancies ($5Nb^{\cdot\cdot\cdot\cdot}_{Li} + 4V'''''_{Nb}$) [27,28], or (iii) niobium antisites and lithium vacancies ($Nb^{\cdot\cdot\cdot\cdot}_{Li} + 4V'_{Li}$) [29-31]. The latter model is now widely accepted in the community [32,33]. From NMR measurements, Yatsenko *et al.* have proposed a defect structure composed of a niobium antisite ($Nb^{\cdot\cdot\cdot\cdot}_{Li}$) surrounded by three lithium vacancies ($V'_{Li}$) in the nearest neighbourhood, plus one independent lithium vacancy ($V'_{Li}$) along the polar z-axis [34,35]. This defect structure is certainly polar and Kim *et al.* have shown that it can explain the presence of an internal field in lithium niobate [36]. The dark conductivity was shown to depend on the [Li]/[Nb] ratio [37].

The aim of our study is to understand the structure of defects at domain walls in lithium niobate for different concentration of magnesium dopants. Crystals under study were commercially available congruent lithium niobate doped with different amount of magnesium: 0 mol% Mg-doped from *CrysTec GmbH*; 1, 2 and 3 mol% from *Crystal Technology*, 5 mol% from *Yamaju Ceramics* and 7 mol% from *Union Optics*. The percentages refer to the concentration range of magnesium with respect to the congruent melt concentration. Samples were cut into 0.5 mm-thick pieces, with short edges along the *c*-direction.

## 2. Results and discussion

Raman-scattering experiments were performed on a Renishaw spectrometer with a low frequency cut-off at 100 cm$^{-1}$. The exciting laser line was 532 nm with a power of 40 mW. The Raman-laser was focused on the sample with an objective with numerical aperture 0.85 providing a theoretical spot size of 760 nm. Raman spectra were collected in parallel $z(xx)\bar{z}$ polarization measurement geometry for which only *E(TO)* and *A₁(LO)* modes are expected, and in crossed $z(yx)\bar{z}$ polarization measurement geometry for which only *A₁(TO)* modes are expected.

Figure 1(a) shows a typical spectrum in $z(xx)\bar{z}$ configuration where 13 modes are assigned accordingly to the work of Margueron *et al.* [38]. The band around 730 cm$^{-1}$ is attributed to a two-phonon scattering process [39]. Detailed analysis is performed on *E(TO₁)*, *E(TO₈)* and *A₁(LO₄)* modes because they are intense and well separated from each-others. All bands are fitted with symmetric Lorentzian profiles using the program *Peak-o-mat* based on *Python*. The vibrational pattern for those three modes are known from first-principle calculations [40-42]: *A₁(LO₄)* involves in-phase vibrations of lithium and niobium ions while they vibrate out of phase for the *E(TO₁)* and *E(TO₈)* modes.

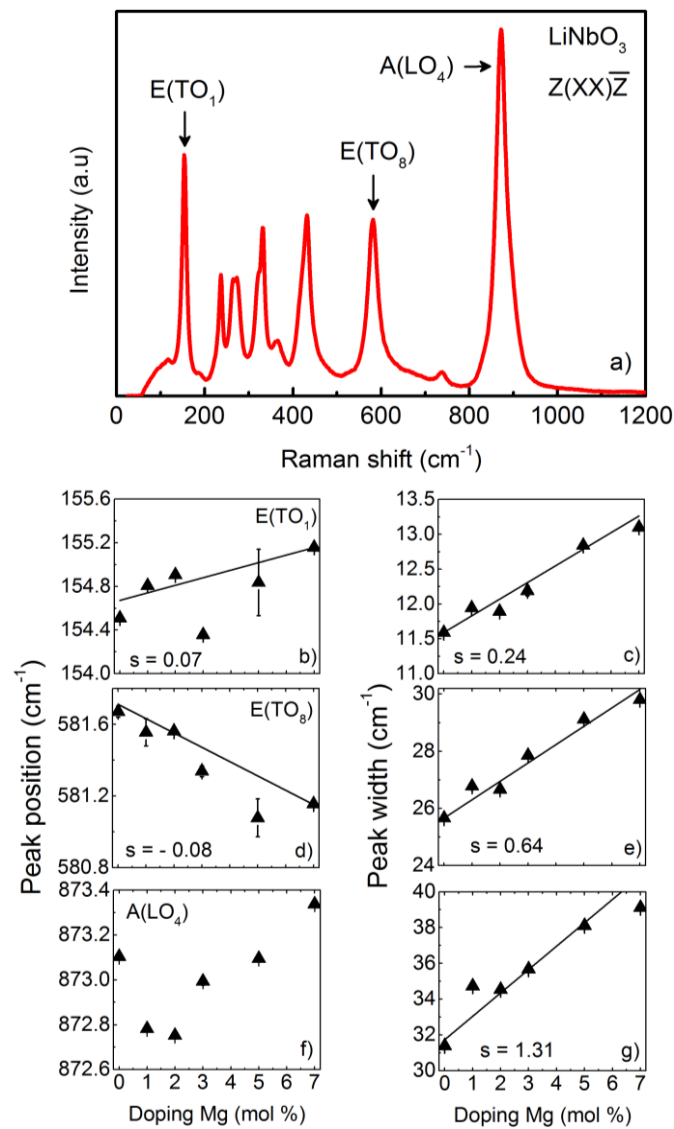

**Figure 1** (a) Typical Raman spectrum of congruent lithium niobate in $z(xx)\bar{z}$ configuration. The other panels describe spectral evolutions with increasing Mg doping levels: (b) frequency and (c) FWHM of *E(TO₁)* modes (d) frequency and (e) FWHM of *E(TO₈)* modes. (f) frequency and (g) FWHM of *A₁(LO₄)*. Black lines are linear fits to the data. Error bars represent the standard deviation calculated from two successive measurements. All spectra were acquired in $z(xx)\bar{z}$ configuration, after the annealing procedure.

Figures 1(b),(d),(f) and Fig. 1(c),(e) and (g) show the evolution of the peak frequency and FWHM, respectively, for different amount of magnesium. Spectra measured in $z(xx)\bar{z}$ and $z(yx)\bar{z}$ configurations revealed the same trend. With increasing magnesium content, the FWHM of all modes increases linearly, which is consistent with an increase of the disorder in the sample due to the incorporation of magnesium. We also note that with increasing magnesium content, the frequency of *E(TO₁)* increases while the frequency of *E(TO₈)* decreases. The change of frequency of *A₁(LO₄)* is not linear. At first sight, no obvious correlation appears, for these three particular modes, between the changes in the Raman modes and the changes in the defect structures with increasing magnesium content.

Ferroelectric domains were then created by applying high voltage pulses through liquid electrodes [43] resulting in a random pattern of hexagonal ferroelectric domains.

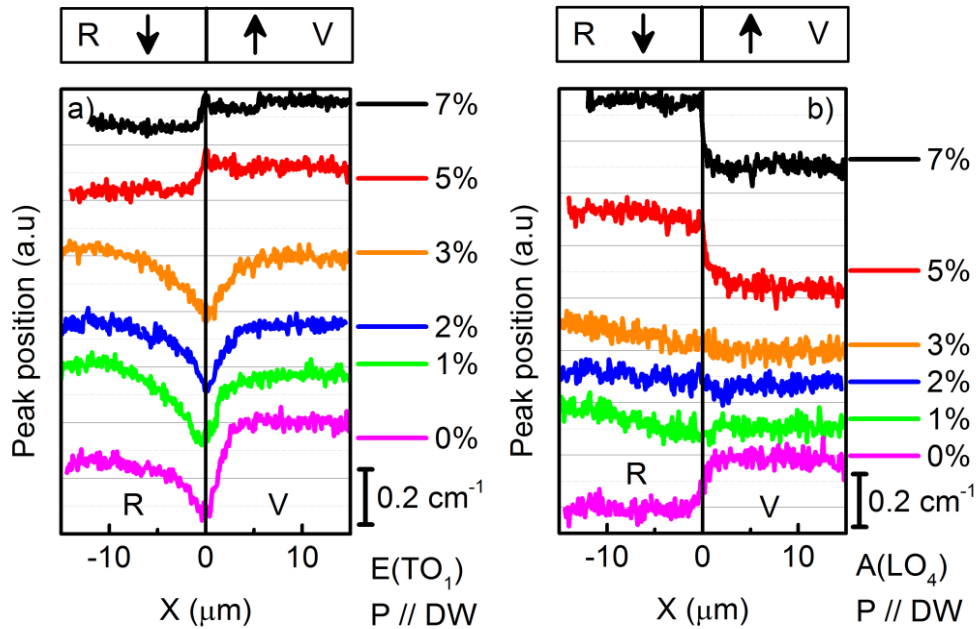

**Figure 2** Peak position of (a) *E(TO₁)* and (b) *A₁(LO₄)* for line-scan from reversed (R) to virgin (V) domains for different amount of magnesium before annealing. The data are vertically shifted for clarity and the scale bar gives the amplitude of the frequency shift.

Figure 2(a) and (b) shows the evolution of the frequency of *E(TO₁)* and *A₁(LO₄)* modes for line-scans with 0.1 µm steps across a domain wall, in $z(xx)\bar{z}$ configuration, with the laser polarisation parallel to the domain wall. Two contrasts are visible: a change of frequency between domains – delta varies between 0.05 and 0.3 cm⁻¹ – and a change of frequency at the domain wall.

We focus first on the contrast between virgin and reversed domains. In the undoped sample, the frequency of *E(TO₁)*, *E(TO₈)* and *A₁(LO₄)* is lower in reversed domains (R) (in our measurement the –z face), consistent with previous observations [44,45]. The frequency shifts have been explained by the presence of polar defects and the internal field they create; this is notably based on the observations that: (i) frequency shifts are larger for the congruent composition than for the stoichiometric composition [44], and (ii) that the frequency shift and the internal field exhibit the same time dependence upon domain reversal [45].

As shown in figure 2, the frequency shifts of the Raman peaks do not change monotonously with Mg doping. In 1%, 2% and 3% Mg-doped lithium niobate the frequency of all modes is slightly higher in reversed domains. On the other hand, in 5% and 7% Mg-doped lithium niobate, the frequency of *E(TO₁)* and *E(TO₈)* is decreased while the frequency of *A₁(LO₄)* is increased in reversed domains. It is known from measurements of the transient current during switching that Mg-doped lithium niobate presents an internal field [46]. By analogy with undoped lithium niobate, we attribute the frequency shifts between domains in Mg-doped lithium niobate to this internal field. The non-monotonous behaviour of the internal field could be understood by a change in the structure of polar defects: the studies of incorporation of Mg$^{2+}$ ions into LiNbO$_3$ have shown that Mg ions first replace $Nb_{Li}^{\cdots}$ for MgO concentrations lower than 4.5-5.3 mol%, and then replace Li and Nb ions on their own sites [27,47-52]. Therefore, the different Raman contrast that we observe for samples doped with more than 5 mol% of magnesium can be explained by a change in the internal field due to (1) the disappearance of all $Nb_{Li}^{\cdots}$, (2) the appearance of $Mg_{Nb}'''$.

The contrast between domains is known to be sensitive to annealing: multiphoton photoluminescence measurements performed on samples doped with between 0% and 7% of magnesium have shown a contrast between virgin and reversed domains only in samples doped with more than 5 mol% of magnesium. Upon annealing above 100°C the contrast obeyed an exponential decay with an activation energy of 1 eV, typical of lithium ions mobility [53]. In 5 mol% Mg-doped lithium niobate, it has been shown that annealing around 200°C highly reduces the internal field [46], probably as a result of reorganization of polar defects [54]. In order to confirm the role of the internal field on the Raman contrast, and focus on the domain wall signature, we annealed our samples at 200°C for 8h under oxygen. As shown in Fig. 3(a) for the *E(TO$_1$)* mode, the contrast between domains indeed vanishes after annealing, but a change of frequency is still observed at domain walls. The results for all three modes are summarized in Fig. 3(b). The frequency shift for *E(TO$_1$)* and *E(TO$_8$)* at the domain wall is negative below 5% of magnesium but positive above 5%, while it is always positive for *A$_1$(LO$_4$)* and nearly independent on the amount of magnesium. It is worth noting that in $z(xx)\bar{z}$ geometry the frequency shift of *E(TO$_8$)* is always positive; this is a side effect due to the large increase of intensity on its shoulder (see G. Stone *et al.* for details [20]) and it is not considered further as a "real" shift. The spatial extension of the frequency shifts varies between 0.6 and 6 μm, depending on the amount of magnesium. For samples doped with more than 5 mol% of magnesium, the spatial extension is always below 1.8 μm.

Frequency shifts of Raman modes at domain walls in lithium niobate have already been reported and interpreted (i) as an effect of the strong electric and associated strain field accompanying smooth polarization reversal at the domain wall [16,17,23] or (ii) as a change in the defect cluster concentration and configurations at the domain wall [20,22]. Specifically, for undoped lithium niobate, our observation of decreasing frequency of *E(TO$_1$)* and *E(TO$_8$)* at the domain wall while increasing frequency of *A$_1$(LO$_4$)* is consistent with previous measurements, which were interpreted as an effect of defect clusters, involving $Nb_{Li}^{\cdots}$ [22].

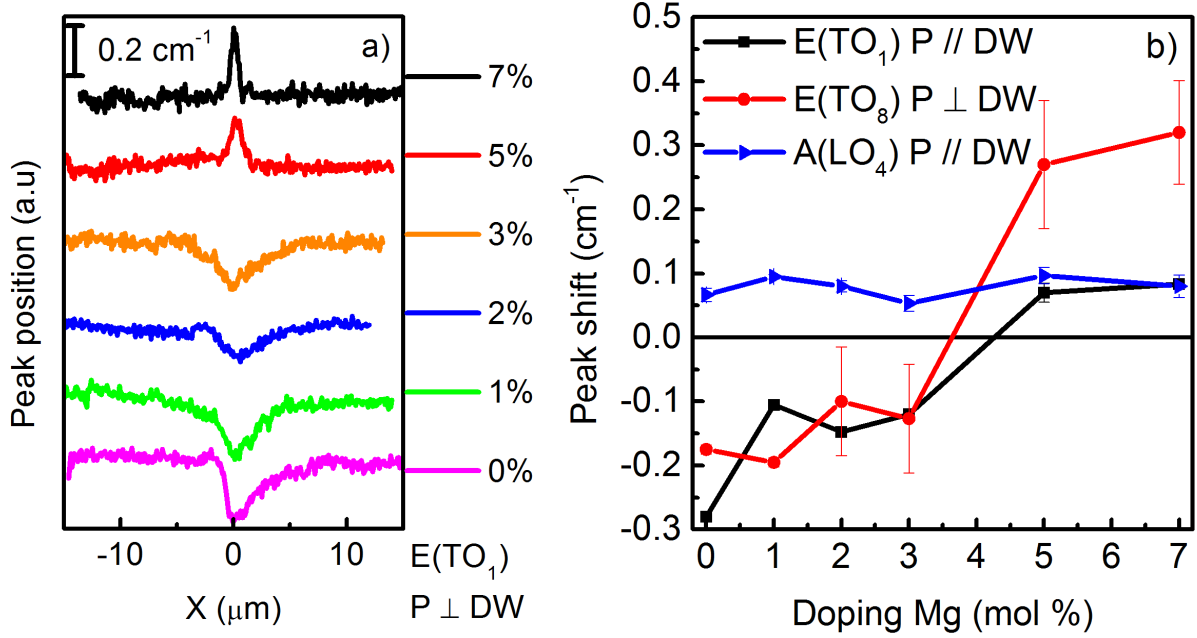

**Figure 3** (a) Peak position of $E(TO_1)$ across a domain wall after annealing. (b) Peak shifts at the domain wall. Error-bars represent the standard deviation calculated from measurements on three different domain walls.

Here, the evolution of the Raman modes for the whole series of Mg concentrations and before and after annealing sheds more light on the origin of the frequency shifts. First, it is clear that an intrinsic electric field alone cannot account for all the observations since it would not explain the doping-dependence of the frequency shifts of the E modes. A reduction of the depolarization field at domain walls due to a better bulk screening could not explain the change of sign of peak shifts. Conversely, a simple picture where the dopant would segregate at the walls would not be consistent with the observations, notably because the FWHM of $E(TO_8)$ and $A_1(LO_4)$ does not change at the domain wall, whereas they vary strongly in the bulk with variations of Mg content. Based on this, we propose that both intrinsic fields and defects have to be taken into account.

On the other hand, the correlation between the frequency shifts for the *E* modes and the change in defect structure for high Mg-doping points to the role of defects, and changes in defect structure. For samples doped with more than 5 mol% of magnesium, the new behaviour observed at domain walls is interpreted as a consequence of the change of configuration of polar defects described earlier. However, detailed examination shows that the frequency pattern at the domain wall differs from the signature of polar defects as observed in the bulk, notably because the doping-dependence is different, as seen by comparing figures 2 and 3. This can be understood in two ways: (1) the frequency shifts observed involve a combination of effects resulting from the defect structure and the electric (strain) field in the vicinity of the wall or (2) new defect structures have been stabilized at domain walls. Since previous measurements based on defect-luminescence microscopy have shown that the defect structure remains similar at the domain wall, the second hypothesis is less plausible [55] [56]. In both cases, it is important to remember that the contrast at domain walls still exists after annealing while it is no longer present in the bulk, meaning that defects have been stabilized at the domain walls but annealed out of the bulk.

## 3. Conclusion

In summary, we reported the evolution of the $E(TO_1)$, $E(TO_8)$ and $A_1(LO_4)$ Raman modes when increasing the amount of magnesium in congruent lithium niobate. We have observed contrasts between virgin and reversed domains, consistent with the literature and identified new frequency shifts of the Raman modes at the domain walls. We have shown differences between the signatures of polar defects in the bulk and at the domain walls, which appear as spaces where defects can be stabilized.

## 4. Acknowledgments

This work was supported by a Pearl grant of the National Research Fund, Luxembourg (FNR/P12/4853155/Kreisel).

**References**

[1] T. J. Yang, V. Gopalan, P. J. Swart, and U. Mohideen, Phys. Rev. Lett. **82**, 4106 (1999).
[2] A. K. Tagantsev, I. Stolichnov, E. L. Colla, and N. Setter, J. Appl. Phys. **90**, 1387 (2001).
[3] J. F. Scott and M. Dawber, Appl. Phys. Lett. **76**, 3801 (2000).
[4] G. Catalan, J. Seidel, R. Ramesh, and J. F. Scott, Rev. Mod. Phys. **84**, 119 (2012).
[5] E. Salje and H. Zhang, Phase Transitions **82**, 452 (2009).
[6] J. Seidel, P. Maksymovych, Y. Batra, A. Katan, S.-Y. Yang, Q. He, A. P. Baddorf, S. V. Kalinin, C.-H. Yang, J.-C. Yang, Y.-H. Chu, E. K. H. Salje, H. Wormeester, M. Salmeron, and R. Ramesh, Phys. Rev. Lett. **105**, 197603 (2010).
[7] S. Farokhipoor and B. Noheda, Phys. Rev. Lett. **107**, 127601 (2011).
[8] O. Diéguez, P. Aguado-Puente, J. Junquera, and J. Íñiguez, Phys. Rev. B **87**, 024102 (2013).
[9] J. Guyonnet, I. Gaponenko, S. Gariglio, and P. Paruch, Adv. Mater. **23**, 5377 (2011).
[10] M. Schröder, A. Haußmann, A. Thiessen, E. Soergel, T. Woike, and L. M. Eng, Adv. Funct. Mater. **22**, 3936 (2012).
[11] M. Schröder, X. Chen, A. Haußmann, A. Thiessen, J. Poppe, D. A. Bonnell, and L. M. Eng, Mater. Res. Express **1**, 035012 (2014).
[12] T. J. Yang, V. Gopalan, P. Swart, and U. Mohideen, J. Phys. Chem. Solids **61**, 275 (2000).
[13] S. Kim, V. Gopalan, and B. Steiner, Appl. Phys. Lett. **77**, 2051 (2000).
[14] T. Jach, S. Kim, V. Gopalan, S. Durbin, and D. Bright, Phys. Rev. B **69**, 064113 (2004).
[15] H. Taniguchi, Y. Fujii, and M. Itoh, J. Ceram. Soc. Japan **121**, 579 (2013).
[16] P. Capek, G. Stone, V. Dierolf, C. Althouse, and V. Gopalan, Phys. Status Solidi **4**, 830 (2007).
[17] P. S. Zelenovskiy, V. Y. Shur, P. Bourson, M. D. Fontana, D. K. Kuznetsov, and E. A. Mingaliev, Ferroelectrics **398**, 34 (2010).
[18] P. S. Zelenovskiy, M. D. Fontana, V. Y. Shur, P. Bourson, and D. K. Kuznetsov, Appl. Phys. A **99**, 741 (2010).
[19] M. D. Fontana, R. Hammoum, P. Bourson, S. Margueron, and V. Y. Shur, Ferroelectrics **373**, 26 (2008).
[20] G. Stone and V. Dierolf, Opt. Lett. **37**, 1032 (2012).
[21] R. Hammoum, M. D. Fontana, P. Bourson, and V. Y. Shur, Appl. Phys. A **91**, 65 (2007).

[22] G. Stone, D. Lee, H. Xu, S. R. Phillpot, and V. Dierolf, Appl. Phys. Lett. **102**, 042905 (2013).
[23] Y. Kong, J. Xu, B. Li, S. Chen, Z. Huang, L. Zhang, S. Liu, W. Yan, H. Liu, X. Xie, L. Shi, X. Li, and G. Zhang, Opt. Mater. (Amst). **27**, 471 (2004).
[24] L. Mateos, L. E. Bausá, and M. O. Ramírez, Opt. Mater. Express **4**, 1077 (2014).
[25] A. Boes, V. Sivan, G. Ren, D. Yudistira, S. Mailis, E. Soergel, and A. Mitchell, Appl. Phys. Lett. **107**, 022901 (2015).
[26] A. M. Prokhorov and Y. S. Kuz'minov, *Physics and Chemistry of Crystalline Lithium Niobate* (CRC Press, Bristol/New York, 1990).
[27] O. F. Schirmer, O. Thiemann, and M. Wöhlecke, J. Phys. Chem. Solids **52**, 185 (1991).
[28] S. C. Abrahams and P. Marsh, Acta Crystallogr. Sect. B Struct. Sci. **42**, 61 (1986).
[29] P. Lerner, C. Legras, and J. P. Dumas, J. Cryst. Growth **3-4**, 231 (1968).
[30] N. Zotov, H. Boysen, F. Frey, T. Metzger, and E. Born, J. Phys. Chem. Solids **55**, 145 (1994).
[31] N. Zotov, F. Frey, H. Boysen, H. Lehnert, A. Hornsteiner, B. Strauss, R. Sonntag, H. M. Mayer, F. Güthoff, and D. Hohlwein, Acta Crystallogr. Sect. B Struct. Sci. **51**, 961 (1995).
[32] V. Gopalan, V. Dierolf, and D. A. Scrymgeour, Annu. Rev. Mater. Res. **37**, 449 (2007).
[33] Y. Li, W. G. Schmidt, and S. Sanna, Phys. Rev. B **91**, 174106 (2015).
[34] A. V. Yatsenko, E. N. Ivanova, and N. A. Sergeev, Phys. B Condens. Matter **240**, 254 (1997).
[35] A. V. Yatsenko, H. M. Ivanova-Maksimova, and N. A. Sergeev, Phys. B Condens. Matter **254**, 256 (1998).
[36] S. Kim, V. Gopalan, K. Kitamura, and Y. Furukawa, J. Appl. Phys. **90**, 2949 (2001).
[37] T. Volk and M. Wöhlecke, *Lithium Niobate* (Springer Berlin Heidelberg, Berlin, Heidelberg, 2008).
[38] S. Margueron, A. Bartasyte, A. M. Glazer, E. Simon, J. Hlinka, I. Gregora, and J. Gleize, J. Appl. Phys. **111**, 104105 (2012).
[39] A. Ridah, M. D. Fontana, and P. Bourson, Phys. Rev. B **56**, 5967 (1997).
[40] P. Hermet, M. Veithen, and P. Ghosez, J. Phys. Condens. Matter **19**, 456202 (2007).
[41] P. Ghosez (*private communication*).
[42] V. Caciuc, A. V. Postnikov, and G. Borstel, Phys. Rev. B **61**, 8806 (2000).
[43] E. Soergel, Appl. Phys. B **81**, 729 (2005).
[44] G. Stone, B. Knorr, V. Gopalan, and V. Dierolf, Phys. Rev. B **84**, 134303 (2011).
[45] J. G. Scott, S. Mailis, C. L. Sones, and R. W. Eason, Appl. Phys. A Mater. Sci. Process. **79**, 691 (2004).
[46] J. H. Yao, Y. H. Chen, B. X. Yan, H. L. Deng, Y. F. Kong, S. L. Chen, J. J. Xu, and G. Y. Zhang, Phys. B Condens. Matter **352**, 294 (2004).
[47] A. V. Yatsenko, S. V. Yevdokimov, D. Y. Sugak, and I. M. Solskii, Funct. Mater. **21**, 31 (2014).
[48] A. V. Yatsenko, S. V. Yevdokimov, D. Y. Sugak, and I. M. Solskii, Acta Phys. Pol. A **117**, 166 (2010).
[49] F. Abdi, M. Aillerie, P. Bourson, and M. D. Fontana, J. Appl. Phys. **106**, 033519 (2009).
[50] B. C. Grabmaier, W. Wersing, and W. Koestler, J. Cryst. Growth **110**, 339 (1991).
[51] Q. R. Zhang and X. Q. Feng, Phys. Rev. B **43**, 12019 (1991).
[52] Y. Chen, W. Yan, J. Guo, S. Chen, G. Zhang, and Z. Xia, Appl. Phys. Lett. **87**, 212904 (2005).

[53] P. Reichenbach, T. Kämpfe, A. Thiessen, A. Haußmann, T. Woike, and L. M. Eng, Appl. Phys. Lett. **105**, 122906 (2014).
[54] N. Meyer, G. F. Nataf, and T. Granzow, J. Appl. Phys. **116**, 244102 (2014).
[55] V. Dierolf, C. Sandmann, S. Kim, V. Gopalan, and K. Polgar, J. Appl. Phys. **93**, 2295 (2003).
[56] V. Dierolf and C. Sandmann, J. Lumin. **125**, 67 (2007).